\documentclass{article}

\usepackage{arxiv}

\usepackage[utf8]{inputenc} % allow utf-8 input
\usepackage[T1]{fontenc}    % use 8-bit T1 fonts
\usepackage{hyperref}       % hyperlinks
\usepackage{url}            % simple URL typesetting
\usepackage{booktabs}       % professional-quality tables
\usepackage{amsfonts}       % blackboard math symbols
\usepackage{nicefrac}       % compact symbols for 1/2, etc.
\usepackage{microtype}      % microtypography
\usepackage{lipsum}		% Can be removed after putting your text content
\usepackage{graphicx}
\usepackage{doi}
\usepackage{amsthm}
\theoremstyle{plain}% Theorem-like structures provided by amsthm.sty
\newtheorem{theorem}{Theorem}

\newtheorem{proposition}[theorem]{Proposition}

\theoremstyle{definition}
\newtheorem{definition}{Definition}

\theoremstyle{remark}
\newtheorem{remark}{Remark}

\title{Natural Gradient in Evolutionary Games}

\author{Vladimir Ja\' cimovi\' c \\
	Faculty of Natural Sciences and Mathematics\\
	University of Montenegro\\ 
	Cetinjski put bb., 81000 Podgorica\\ 
	Montenegro\\
	\texttt{vladimirj@ucg.ac.me} \\
}

\renewcommand{\shorttitle}{Natural Gradient in Evolutionary Games}

\hypersetup{
pdftitle={Natural Gradient in Evolutionary Games},
pdfsubject={-},
pdfauthor={Vladimir Ja\' cimovi\' c},
pdfkeywords={replicator equations, Fisher information metric, natural evolution strategies, information-geometric optimization},
}

\begin{document}
\maketitle

\begin{abstract}
We study evolutionary games with a continuous trait space in which replicator dynamics are restricted to the manifold of multidimensional Gaussian distributions. We demonstrate that the replicator equations are natural gradient flow for maximization of the mean fitness.

Our findings extend previous results on information-geometric aspects of evolutionary games with a finite strategy set.

Throughout the paper we exploit the information-geometric approach and the relation between evolutionary dynamics and Natural Evolution Strategies, the concept that has been developed within the framework of black-box optimization. This relation sheds a new light on the replicator dynamics as a compromise between maximization of the mean fitness and preservation of diversity in the population.
\end{abstract}

% keywords can be removed

\section{Introduction}\label{sec:1}

Simple games, that serve as illustrative examples in Evolutionary Game Theory (EGT), belong to the class of {\it evolutionary games with a finite strategy set}. In such a setup, one investigates a {\it population} that consists of $n$ {\it species}. Denote the set of species by $A = \{A_1,\dots,A_n\}$ and consider the set ${\cal P}(A)$ of all probability distributions over $A$. Set ${\cal P}(A)$ can be identified with the unit simplex in $n$-dimensional vector space:
$$
{\cal P}(A) \simeq \Delta^n = \{ p = (p_1 \dots p_n)^T \, : \, p_1 + \cdots + p_n = 1, \; p_i \geq 0 \; \; i=1,\dots,n \}.
$$
Probability $p_i = P(A_i)$ can be conceived as a portion of species $A_i$ in the population.

Elements of ${\cal P}(A)$ are called {\it strategies}. Those strategies that are concentrated at a single species (i.e. probability distributions with $p_i = 1$ for some $i \in \{1,\dots,n\}$) are named {\it pure strategies}. Hence, there are precisely $n$ pure strategies that correspond to $n$ species. Strategies that are note pure are called {\it mixed strategies}.

Furthermore, to each $A_i \in A$ we assign a continuous function $f_i(p) \equiv f_i(p_1,\dots,p_n) : {\mathbb R}^n \to {\mathbb R}$. Functions $f_i(p)$ are called {\it fitness functions}. Vector-valued function $f(p) \equiv (f_1(p),\dots,f_n(p)) : {\mathbb R}^n \to {\mathbb R}^n$ defines so-called {\it fitness landscape}. The assumption that the fitness of one species depends on portions of all the others, brings a strategic aspect into interactions between them.

Introduce the mean fitness in the population:
\begin{equation}
\label{mean_fitness}
\langle f(p) \rangle = p \cdot f(p) = \sum \limits_{i=1}^n p_i f_i(p).\end{equation}
Throughout this paper the notion $x \cdot y$ stands for the inner product of vectors $x$ and $y$ in the vector space. Sometimes we will also use the notation $x^T y$.

We assume that portions $p_i$ evolve with the time according to the following ODE's
\begin{equation}
\label{replicator}
\dot p_i = p_i \left( f_i(p) - \langle f(p) \rangle \right), \quad i = 1,\dots,n.
\end{equation}
Equations (\ref{replicator}) are well known as {\it replicator equations}. They have a simple justification. We say that the state of population at a moment $t$ is given by vector $p(t) = (p_1(t),\dots,p_n(t))$. Then equations (\ref{replicator}) claim that if for a given state of population, species $A_i$ has higher fitness than the mean fitness, its portion $p_i(t)$ will increase.

Foundations of EGT have been laid in 1970's by Maynard Smith (\cite{Maynard Smith}), who applied game-theoretic paradigms to the Darwinian theory of evolution. Hence, the terminology inherited from the biological context (species, fitness, etc.) is commonly used in the literature. As EGT emerged into a prominent branch of Game Theory, it has been recognized that its models and paradigms present a significant interest for social sciences and Philosophy. Indeed, the evolution is not exclusively biological concept; one can talk, for instance, about cultural, or behavioral evolution. Taking into account such a wide scope of interpretations, elements of the set $A$ do not necessarily represent biological species. Depending on the field of applications, $A_i$ can correspond to political views, lifestyle habits, fashion preferences, or moral choices. Such a variety of interpretations brought alternative terminologies into EGT. In some cases it is more appropriate to talk about {\it actions} instead of species, {\it payoff} instead of fitness; in these cases (\ref{mean_fitness}) represents an {\it expected payoff}.

Regardless of applications and interpretations, the abstract mathematical framework of EGT remains the same. It is easy to check that $\Delta^n$ is an invariant set for the dynamics (\ref{replicator}). Hence, replicator equations describe the evolution of strategies, that is - of probability distributions over a finite set. Distributions from ${\cal P}(A)$ are usually called {\it categorical distributions}.

In addition to evolutionary games with a finite strategy set, one can also conceive a situation where the set of actions (that is - of pure strategies) is a continuous subset of ${\mathbb R}^n$. Then mixed strategies are absolutely continuous probability measures on ${\mathbb R}^n$. In such a setup, evolutionary dynamics generate a flow on a certain family of probability measures on ${\mathbb R}^n$. Such games are named {\it evolutionary games with a continuous trait space}. \footnote{Games with a continuous space of pure strategies allow for a  biological interpretation in which pure strategies correspond to individual traits of a certain species. Such an interpretation stands behind the expression "continuous trait space", see for instance \cite{CHR,Karev}. On the other hand, the term "game with a finite strategy set" is a bit imprecise, since it means that the set of {\bf pure} strategies is finite. (The set of all strategies is continuous in any case.) Nevertheless, the term "games with a continuous strategy space" is commonly used in the literature. Alternatively, some researchers (\cite{MZ}) preferred to talk about evolutionary games with {\it finite (or continuous) action sets}. This terminology is inspired by various non-biological interpretations of evolutionary games.}

Mathematical framework of EGT makes it possible to study evolutionary games as flows on families of probability distributions. Such an approach relies on results and paradigms of Information Geometry (IG). Below, we will explain information-geometric approach to evolutionary games and exploit it throughout the present paper.

When investigating evolutionary games with a finite strategy set, one should take into account metric properties of ${\cal P}(A)$. It is not difficult to notice that the standard Euclidean metric is an inappropriate measure of the distance between two categorical distributions. Instead, IG proposes a general way to introduce a metric on families of probability distributions, thus turning them into Riemannian manifolds. Metric further defines a gradient flow on the manifold and replicator dynamics can be studied from this point of view. In the case of games with a finite strategy set, this has been recognized by Marc Harper and led to novel insights into geometric and information-theoretic aspects of evolutionary dynamics, \cite{Harper1,Harper2}. In Section \ref{sec:2} we briefly recall Harper's results that serve as a starting point for further generalizations.

Manifold ${\cal P}(A)$ of categorical distributions is just one particular example of the so-called {\it statistical manifold}. In Section \ref{sec:3} we briefly expose general geometric approach to families of probability distributions and introduce the notions of {\it Fisher information metric} and {\it natural gradient}. From this point of view, Harper's results about the evolutionary dynamics on the manifold of categorical distribution might turn out to be just the tip of the iceberg. This suggests that IG provides a universal theoretical background for study of a broad class of evolutionary games. One can treat evolutionary dynamics as gradient flows on statistical manifolds. Although information-geometric concepts are widely used in Mathematics and Computer Science, the most comprehensive theoretic approach to gradient flows on statistical manifolds has been developed for purposes of black-box optimization. Black-box optimization (including evolutionary algorithms and stochastic search methods) relied for decades on various heuristics without a coherent theoretical justification. However, this direction of research evolved into a general framework that encompasses some of previously known stochastic search algorithms and provides a solid theoretical background for them. The black-box optimization community adopted the terms {\it Natural Evolution Strategies} (NES) and {\it Information-Geometric Optimization} (IGO) for gradient flows on statistical manifolds, \cite{WSGSPS,WSPS,HA,OAAH,Glasmachers,ANOK}. NES and IGO offer an abstract conceptual approach which is valid for any statistical manifold. Section \ref{sec:4} contains an explanation of these concepts with a view on their relevance for EGT.

In Section \ref{sec:5}, following papers \cite{CHR,Ruijgrok^2}, we introduce the general setup of games with a continuous trait space: fitness landscape, replicator equations, etc.

A notable example of statistical manifold is obtained by endowing the family of Gaussian distributions with the Fisher information metric.
In Section \ref{sec:6} we study restriction of the evolutionary dynamics to this particular statistical manifold (we denote it by ${\cal N}(a,C)$). We demonstrate that the replicator equations are natural gradient flow on ${\cal N}(a,C)$ for maximization of the mean fitness. This assertion is an extension of the Harper's result about gradient flows on the statistical manifold ${\cal P}(A)$.

In Section \ref{sec:7} we apply the NES and IGO approaches to evolutionary games with Gaussian mixed strategies. This demonstrates the relation between recent advances in black-box optimization and EGT. Our study unveils the global objective that the population as a whole tends to achieve. These findings also support an interpretation of evolutionary dynamics as a multi-agent learning algorithm.

Analysis of asymptotic properties of replicator dynamics suggests that the corresponding optimization algorithm exhibits a slow convergence. In Section \ref{sec:8} we borrow ideas from the black-box optimization in order to investigate mechanisms that accelerate evolutionary dynamics (and hence the corresponding algorithms). By introducing these mechanisms into evolutionary dynamics, one can design algorithms with significantly higher convergence rate, at the expense of diversity. We also briefly point out some biological interpretations of these mechanisms that accelerate the evolution.

Finally, Section \ref{sec:9} contains some concluding remarks, a brief discussion on various applications and an outlook into future research directions.

The main result is reported in Section \ref{sec:6}. Novel insights are also exposed in sections \ref{sec:7} and \ref{sec:8}. Sections \ref{sec:1}-\ref{sec:5} serve as a necessary introduction and are included for the sake of completeness of the exposition. The present paper is mostly inspired by papers \cite{CHR},\cite{Beyer} and \cite{Harper1}. On one side, Cressman at al. in \cite{CHR} study the restriction of replicator dynamics to the family of Gaussian distributions and derive ODE's for the mean vector and covariance matrix. On the other side, Beyer in \cite{Beyer} investigates the natural gradient policy for optimization of linear-quadratic function. Adaptation of his analysis to the setup of an evolutionary game, yields precisely the system of ODE's derived in \cite{CHR}. Therefore, we conclude that evolutionary dynamics on the manifold ${\cal N}(a,C)$ generate natural gradient flow for the mean fitness.

\section{Natural gradient dynamics in games a with finite strategy set}\label{sec:2}

In order to study information-geometric aspects of evolutionary dynamics, we start by introducing metric on the family of categorical distributions. As already mentioned, this family can be identified with the unit simplex.

Each point of the simplex satisfies $p_1 + \cdots + p_n = 1$, so the tangent space at any interior point of $\Delta^n$ is $(n-1)$-dimensional vector space consisting of vectors $v = (v_1 \cdots v_n)^T$ that satisfy $v_1 + \cdots + v_n = 0$. Orthogonal complement of this space is the line with direction vector ${\bf 1} = (1 \; \cdots \; 1)^T$. We introduce the following Riemannian metric tensor on the interior of $\Delta^n$
\begin{equation}
\label{metric}
g(\eta,\nu) = \sum \limits_{i=1}^n \frac{1}{p_i} \eta_i \nu_i,
\end{equation}
where $p$ is a point in the interior of $\Delta^n$ and $\eta$ and $\nu$ are vectors from the tangent space at $p$.

Metric $g$ turns the unit simplex $\Delta^n$ (and, hence, the family of categorical distributions ${\cal P}(A)$) into a Riemannian manifold.

\begin{remark}
In EGT (\ref{metric}) is commonly referred to as {\it Shahshahani metric}, because it has been introduced into Game Theory in paper \cite{Shahshahani}. Notice that $g$ diverges at the boundary of $\Delta^n$ (since on the boundary there exists $i$, such that $p_i=0$), so the definition is valid only in the interior.
\end{remark}

\begin{theorem} \cite{HSS,Harper1}
If the system of differential equations $\dot p_i = f_i(p)$ defines a Euclidean gradient flow with $f_i(p) = \frac{\partial V}{\partial p_i}$, then replicator equations (\ref{replicator}) define a gradient flow with respect to the Shahshahani metric.
\end{theorem}

\begin{remark}
In the literature on EGT it is often assumed that fitness functions are linear:
$$
f_i(p) = \sum \limits_{j=1}^n a_{ij} p_j,
\mbox{   for some coefficients } a_{ij}.
$$
This assumption yields the linear fitness landscape $f(p) = A p$ for matrix $A = \{a_{ij}\}$ which is called fitness (or payoff) matrix.

For linear fitness functions Theorem 1 becomes even more transparent. Indeed, Euclidean gradient is $Ap$ and Shahshahani potential turns out to be the mean fitness: $\frac{1}{2} p \cdot f(p) = \frac{1}{2} p \cdot A p$.
\end{remark}

Given two probability distributions $p, q \in {\cal P}(A)$, the Kullback-Leibler divergence (sometimes also referred to as {\it relative entropy}) between them is defined as:
\begin{equation}
\label{K-L}
I_{KL}(p \, || \, q) = \sum \limits_{i=1}^n q_i \ln \left( \frac{q_i}{p_i}\right).
\end{equation}
K-L divergence is a measure how "distant" (i.e. how easily distinguishable given the random sample) two probability distributions are. It does not define a distance function on ${\cal P}(A)$, since it is not symmetric, i.e. $I_{KL}(p \, || \, q) \neq I_{KL}(q \, || \, p)$. Still, (as any divergence function) (\ref{K-L}) is nonnegative and satisfies $I_{KL}(p \, || \, q) = 0 \iff p \equiv q$.

The Taylor expansion of the K-L divergence up to the second order term along the diagonal $p=q$ yields:
\begin{equation}
\begin{array}{ll}
\label{KL_expansion}
I_{KL}(p \, || \, q) = I_{KL}(p \, || \, q)|_{p=q} + (\nabla I_{KL}(p \, || \, q)|_{p=q}) \cdot (p-q) + \frac{1}{2} (p-q)^T H(p) (p-q) + \cdots = \\
\\
= 0 + 0 + \frac{1}{2} (p-q)^T H(p) (p-q) + \cdots.
\end{array}
\end{equation}
Here, $H(p)$ denotes the Hessian matrix of $I_{KL}$:
$$
H(p) = \left. \left (\frac{\partial^2 I_{KL}}{\partial p_i \partial q_j} \right) \right \vert_{p=q}.
$$
Note that the first order term in (\ref{KL_expansion}) equals zero, because gradient is parallel to the vector ${\bf 1} = (1 \;\cdots \; 1)^T$, so ${\bf 1} \cdot (p-q) = {\bf 1} \cdot p - {\bf 1} \cdot q = 1-1 = 0.$

An easy calculation yields $H(p) = diag \{1/p_1,\dots,1/p_n\}$. Hence, Shahshahani metric (\ref{metric}) is locally defined by the K-L divergence on ${\cal P}(A)$.

Following the pioneering works of Maynard Smith, one of central concepts in EGT is the notion of an {\it evolutionary stable strategy} (ESS). A strategy $\hat p \in {\cal P}(A)$ is said to be ESS if $\hat p \cdot f(p) > p \cdot f(p)$ for all $p$ in some neighborhood of $\hat p$.

\begin{theorem} \cite{Harper1}
\label{KL_Lyapunov}
The strategy $\hat p$ is an interior ESS, if and only if $I_{KL}(\hat p \, || \, p)$ is a Lyapunov function for the dynamics (\ref{replicator}).
\end{theorem}

Theorem 2 can be found in \cite{Harper1}, but its versions have been reported earlier, see \cite{Akin1,HSS}.

Theorems 1 and 2 unveil an information-theoretic background of evolutionary games with a finite strategy set. Evolutionary dynamics appears as a learning process. K-L divergence can be seen as the amount of information left to learn until the population achieves ESS (for a nice and enlighting exposition of the above interpretations we refer to the paper \cite{BP} of Baez and Pollard).

\section{Fisher information metric on statistical manifolds}\label{sec:3}

Let $P$ and $Q$ be absolutely continuous random variables on a measurable space $X \subseteq {\mathbb R}^n$ with densities $p(x)$ and $q(x)$ respectively. The K-L divergence between $P$ and $Q$ is defined as:
$$
I_{KL}(P \, || \, Q) = \int_X p(x) \ln \left( \frac{p(x)}{q(x)} \right) dx.
$$
K-L divergence is not symmetric, but its infinitesimal version is. More precisely, suppose that $P(\theta+\delta \theta)$ and $P(\theta)$ are two random variables that belong to the statistical manifold ${\cal P}$ and $\delta \theta$ is an infinitesimally small variation of the parameter. Then the K-L divergence satisfies $I_{KL}(P(\theta + \delta \theta) \, || \, P(\theta)) = I_{KL}(P(\theta) \, || \, P(\theta + \delta \theta))$ and the Taylor expansion up to the second order term yields:
\begin{equation}
\label{KL_expansion1}
I_{KL}(P(\theta + \delta \theta) \, || \, P(\theta)) = 0 + 0 + \frac{1}{2} \delta \theta^T F(\theta) \delta \theta + o(\delta \theta^T \delta \theta),
\end{equation}
where $F$ is the matrix whose entries are
\begin{equation}
\label{Fisher_metric}
f_{ij} (\theta) = \mathbb{E} \left[ \frac{\partial \log q}{\partial \theta^i} \frac{\partial \log q}{\partial \theta^j} \right].
\end{equation}

\begin{definition}
Matrix $F(\theta)$ with entries (\ref{Fisher_metric}) is called {\it Fisher information matrix}. The corresponding metric is called {\it Fisher information metric} on ${\cal P}$. A family of probability measures, endowed with this metric, is said to be a {\it statistical manifold} with the manifold coordinates $\theta$.
\end{definition}

\begin{remark}
The Fisher information matrix for the family of categorical distributions can easily be evaluated explicitly (see \cite{AF}):
$$
f_{ij} (p)= \mathbb{E} \left[ \frac{\partial \log p}{\partial p^i} \frac{\partial \log p}{\partial p^j} \right] = \sum \limits_{k=1}^n p_k \frac{1}{p_i} \delta_{ik} \frac{1}{p_j} \delta_{jk} = \frac{1}{p_i} \delta_{ij}.
$$
Therefore, the Shahshahani metric (\ref{metric}) is a particular case of the Fisher information metric on the manifold of categorical distributions.
\end{remark}

\section{Natural Evolution Strategies}\label{sec:4}

Consider the continuous optimization problem
$$
f(x) \to \max_{x \in {\mathbb R}^n}.
$$
Following the terminology that is common in the black-box optimization, the objective function $f(x)$ will sometimes be called the "fitness function".

A classical approach to the above maximization problem is the gradient ascent method, or its improvements that use both gradient and the Hessian of $f(x)$. However, in many cases gradients of the objective function are difficult or impossible to evaluate.

An alternative to gradient-based optimization methods is provided by stochastic search algorithms. These methods search for a probability distribution $p(x\, | \, \theta)$ that maximizes mathematical expectation of $f(x)$. In order to design a reasonably simple algorithm, it is assumed that $p(x\, | \, \theta)$ belongs to a certain family of probability distributions ${\cal P}$, that depend on parameter $\theta$. Then stochastic search algorithms perform an iterative update of $\theta$. The objective is to maximize the {\it expected fitness} over statistical manifold ${\cal P}$:
\begin{equation}
\label{exp_fitness}
J(\theta) = {\mathbb E}_\theta [f(x)] = \int f(x) p(x \, | \, \theta) dx \to \max_{\theta}.
\end{equation}

Using the so-called {\it log-likelihood trick} one can write:
$$
\nabla_\theta J(\theta) = \nabla_\theta \int f(x) p(x \, | \, \theta) dx = \int f(x) \nabla_\theta p(x \, | \, \theta) dx = \int f(x) \nabla_\theta p(x \, | \, \theta) \frac{p(x\, | \, \theta)}{p(x \, | \, \theta)} dx =
$$
$$
 \int [f(x) \nabla_\theta \log p(x \, | \, \theta)] p(x\, | \, \theta) dx = {\mathbb E}_\theta [f(x) \nabla_\theta \log p(x \, | \, \theta)].
$$
Hence, the search gradient can be estimated from samples $x_1,\dots,x_m$:
\begin{equation}
\label{grad_est}
\nabla_\theta J(\theta) \approx \frac{1}{m} \sum_{i=1}^m f(x_i) \nabla_\theta \log p(x_i \, | \, \theta).
\end{equation}

Now, one can perform the gradient ascent update of $\theta$, using the gradient estimate (\ref{grad_est})
\begin{equation}
\label{gradient_ascent}
\theta^{(t+\delta t)} = \theta^t + \delta t \cdot \nabla J(\theta^t).
\end{equation}
In the limit of an infinitesimally small step $\delta t \to 0$, this algorithm yields the gradient flow:
\begin{equation}
\label{gradientODE}
\frac{d \theta}{dt} = \nabla J(\theta) \bigg|_{\theta = \theta(t)}.
\end{equation}
However, algorithm (\ref{gradient_ascent}) does not take into account non-Euclidean geometry of the statistical manifold ${\cal P}$. In the above formulas $\nabla J(\theta)$ denotes the vector of partial derivatives of $J$. This vector is gradient of $J$ only if $\theta$ is a vector in the Euclidean space equipped with the standard inner product. If the metric is defined by a positive definite matrix $A$, then the gradient with respect to this metric is $A^{-1} \nabla.$

Recall that we aim to maximize the expected fitness (\ref{exp_fitness}) over a family of probability distributions $p(x \, | \, \theta)$. By introducing metric (\ref{Fisher_metric}) on this family, we obtained an optimization problem over a statistical manifold equipped with the Fisher information metric. The gradient associated with this metric is:
\begin{equation}
\label{nat_grad}
\tilde \nabla_\theta = F^{-1}(\theta) \nabla_\theta,
\end{equation}
where $F$ is the Fisher information matrix. Therefore, the iterative procedure (\ref{gradient_ascent}) is to be modified: $\theta^{(t+\delta t)} = \theta^t + \delta t \cdot F^{-1}(\theta^t) \nabla J(\theta^t)$ and the corresponding gradient flow is given by the following ODE
\begin{equation}
\label{nat_grad_ODE}
\frac{d \theta}{dt} = \tilde \nabla J(\theta) \bigg |_{\theta = \theta(t)} = F^{-1}(\theta) \nabla J(\theta) \bigg |_{\theta = \theta(t)}.
\end{equation}
Following pioneering works of Amari \cite{Amari1,AN} on Information Geometry, $\tilde \nabla$ is called {\it natural gradient}. On the other hand, in the literature on Optimization and Machine Learning $\nabla$ is usually referred to as {\it conventional} or {\it vanilla} gradient. Obviously, the conventional gradient is not very meaningful when optimizing over statistical manifolds.

In order to explore some properties of the natural gradient, consider the following maximization problem:
\begin{eqnarray}
\label{nat_grad_opt}
J(\theta + \delta \theta) - J(\theta) \to \max_{\delta \theta} \nonumber \\
\mbox{  so that    } I_{KL}(p(\theta + \delta \theta) \, || \, p(\theta)) = \varepsilon.
\end{eqnarray}
Hence, we are looking for an update $\delta \theta$ in the direction that yields a maximal increase of the expected fitness, while imposing the constraint on the information gain at each step.

In order to solve (\ref{nat_grad_opt}), expand the objective function:
$$
J(\theta + \delta \theta) - J(\theta) = \nabla J(\theta) \cdot \delta \theta + \cdots
$$
and introduce the Lagrange function (using (\ref{KL_expansion1}) and neglecting higher order terms)
$$
L(\delta \theta,\lambda) = \nabla J(\theta) \cdot \delta \theta + \lambda (\varepsilon - \frac{1}{2} \delta \theta \cdot F(\theta) \delta \theta).
$$
Taking derivatives with respect to $\delta \theta$ and $\lambda$:
\begin{equation}
\label{Lagrange_derivative_theta}
\frac{\partial L}{\partial \delta \theta} = \nabla J(\theta) - \lambda F(\theta) \, \delta \theta
\end{equation}
\begin{equation}
\label{Lagrange_derivative_lambda}
\frac{\partial L}{\partial \lambda} = \varepsilon - \frac{1}{2} \delta \theta \cdot F(\theta) \delta \theta.
\end{equation}
Equating (\ref{Lagrange_derivative_theta}) to zero, one gets $\nabla J(\theta) - \lambda F \delta \theta = 0.$ Solving for $\delta \theta$ yields
$$
\delta \theta = \frac{1}{\lambda} F^{-1}(\theta) \nabla J(\theta).
$$
By substituting $\delta \theta$ into (\ref{Lagrange_derivative_lambda}) and using (\ref{Lagrange_derivative_theta}) one obtains expressions for $\varepsilon$ and $\lambda$. In the limit of infinitesimally small time increment this yields the following ODE:
\begin{equation}
\label{ODE_nat_grad}
\frac{d \theta}{dt} = F^{-1}(\theta) \nabla J(\theta)\bigg|_{\theta = \theta(t)}
\end{equation}
which is precisely the natural gradient flow for maximization of $J(\theta)$.

The above derivation can be summarized in the following

\begin{theorem} (\cite{Beyer,OAAH})
The natural gradient ascent algorithm follows the direction $\delta \theta$ on statistical manifold ${\cal P}$ that achieves a maximal increase in expected fitness $J(\theta)$, for a given K-L divergence between random variables $P(\theta + \delta \theta)$ and $P(\theta)$.
\end{theorem}

\begin{definition}
ODE (\ref{ODE_nat_grad}) is said to be the {\it natural gradient flow}.
\end{definition}

\begin{remark}
NES and IGO offer an abstract and universal approach to evolutionary optimization. In theory, one can choose any family of probability distributions in order to perform stochastic search. However, these methods require evaluation (and inversion) of the Fisher information matrix, which is computationally expensive. Some of the most successful algorithms optimize over ${\cal N}(a,C)$, because it is one of a few statistical manifolds for which there exists an explicit expression for the Fisher information matrix.

One of the most successful methods for the black-box optimization, named CMA ES (Covariance Matrix Adaptation - Evolution Strategies), operates on the family of Gaussian distributions by updating the mean vector and covariance matrix. This method has been proposed in \cite{HO} before the concept of NES was developed. After NES have been introduced, it has been recognized that CMA ES fits in this broad framework, as a notable member of NES, see \cite{ANOK}.
\end{remark}

\begin{remark}
In practice, all stochastic search algorithms use some sort of quantile sampling from distributions $p(x \, | \, \theta)$. This approach, based on NES along with quantile sampling, is named Information-geometric optimization (IGO), and the corresponding continuous-time trajectories obtained from modification of (\ref{ODE_nat_grad}) are called {\it IGO flow}, see \cite[Def. 4]{OAAH}.
\end{remark}

\section{Evolutionary games with a continuous trait space}\label{sec:5}

In this Section we briefly introduce evolutionary games in which strategies are probability measures on a continuous space $S \in {\mathbb R}^n$. Depending on the context, points in $S$ can be interpreted as individual (biological) traits or actions undertaken by players. Strategy of the population at the moment $t$ is given by a probability measure $P(t)$ on a measurable space $(S,{\cal B})$, where ${\cal B}$ is the $\sigma$-algebra of Borel subsets of $S$.
We will also say that the {\it population is in the state} $P(t)$ at the moment $t$. Denote by $\Delta(S)$ the set of all probability measures with respect to $(S,{\cal B})$.

For $A \in {\cal B}$, $P(t)(A) = \int_A P(t) ds$ is the proportion of traits (or actions) belonging to the set $A$ in the set of all traits (actions) at the moment $t$.

The fitness landscape is given by a continuous real-valued function $f : S \times S \to {\mathbb R}^n$.

Then the fitness (expected payoff) of the strategy $Q \in \Delta(S)$ played against population in the state $P$ is
\begin{equation}
\label{exp_payoff}
\pi(Q,P) = \int_S \int_S f(s,y) Q(ds) P(dy).
\end{equation}
A trait $s \in S$ is identified with the delta distribution $\delta_s \in \Delta(S)$ that assigns probability one to $s$. These distributions are pure strategies. Then, from (\ref{exp_payoff}) we have that the fitness of a trait $s$ when the population is in the state $P$ is given by:
$$
\pi(s,P) \equiv \pi(\delta_s,P) = \int_S \int_S f(s,y) \delta_s(ds) P(dy) = \int_S f(s,y) P(dy).
$$
Furthermore, integration over $S$ yields an expected fitness in the population
\begin{equation}
\label{exp_fitness_P}
\langle \pi(P) \rangle \equiv \pi(P,P) = \int_S f(\delta_s,P) P(ds).
\end{equation}
Finally, relative fitness of a trait $s$ against the population in the state $P$ is
$$
\phi(s,P) = \pi(s,P) - \langle \pi(P) \rangle.
$$
The replicator equation reads (see \cite{CHR,Ruijgrok^2}):
\begin{equation}
\label{cont_replicator}
\frac{dP}{dt}(A) = \int_A \phi(s,P) P(ds) = \int_A (\pi(\delta_s) - \pi(P,P)) P(ds).
\end{equation}
Equation (\ref{cont_replicator}) has the same meaning as replicator equations (\ref{replicator}): the probability of traits belonging to the set $A$ increases if they have higher expected fitness than the mean fitness in the population. Hence, under evolutionary dynamics (\ref{cont_replicator}) the probability measure $P(t)$ at each moment $t$ tends to concentrate around those subsets of $S$ in which traits have high fitness.

Observe that (\ref{cont_replicator}) is an infinite-dynamical system on the space $\Delta(S)$ of probability measures. Uniqueness and existence issues for (\ref{cont_replicator}) are discussed in \cite{OR}. Obviously, it is very difficult to study or simulate dynamics (\ref{cont_replicator}). Instead, it makes sense to restrict the dynamics to a certain family of probability measures on $S$, thus studying the replicator dynamics on a particular statistical manifold. To that end, we introduce two assumptions that greatly simplify the situation:

{\bf Assumption 1:} The fitness landscape $f$ is given by a bilinear-quadratic function:
\begin{equation}
\label{quadbilin}
f(s,y) = - s \cdot Q s + s \cdot B y,
\end{equation}
where $Q$ is a positive symmetric $n \times n$ matrix and $B$ is an arbitrary $n \times n$ matrix.

The first term in (\ref{quadbilin}) describes the internal fitness of a trait $s$, while interactions across the population are given by the second term (that is - by matrix $B$).

{\bf Assumption 2:} The initial state of population $P(0) = P_0$ is given by a multivariate Gaussian distribution $N(a,C)$, with mean vector $m \in {\mathbb R}^n$ and positive-definite covariance matrix $C \in {\mathbb R}^{n \times n}$.

It is not difficult to check that quadratic-bilinear fitness guarantees invariance of the family of Gaussian distributions for dynamics (\ref{cont_replicator}). Moreover, from (\ref{cont_replicator}) one can derive explicit ODE's for the mean vector and covariance matrix.

\begin{theorem} \cite{CHR}
Under assumptions 1 and 2 the Gaussian family of measures is forward-invariant for the dynamics (\ref{cont_replicator}). Assume that the initial distribution $P(0) = N(a(0),C(0))$ is Gaussian with the mean vector $m(0)$ and covariance matrix $C(0)$. Then, the solution of (\ref{cont_replicator}) is given by $P(t) = N(a(t),C(t))$, with parameters $a(t)$ and $C(t)$ satisfying the following system
\begin{eqnarray}
\frac{da(t)}{dt} = C(t)(B - 2Q) a(t); \nonumber \\
\frac{dC(t)}{dt} = - 2 C(t) Q C(t). \nonumber
\end{eqnarray}
\end{theorem}

\section{Natural gradient dynamics in evolutionary games with a continuous trait space}\label{sec:6}

In this Section we work under assumptions 1 and 2 from the previous Section, thus restricting the dynamics to the manifold ${\cal N}(a,C)$.

Consider the problem of maximization of the mean fitness as defined in (\ref{exp_fitness_P}):
\begin{equation}
\label{exp_fitness_max}
\langle \pi(P) \rangle = {\mathbb E}_P [f(s,y)] \to \max_{P \in {\cal N}(a,C)}
\end{equation}
In other words, we are looking for the Gaussian probability measure that maximizes the mean fitness (expected payoff) in the population. Maximization problem (\ref{exp_fitness_max}) can be rewritten as:
\begin{equation}
\label{exp_fitness_max1}
J(a,C) = {\mathbb E}_{a,C} [f(s,y)] = {\mathbb E}_{a,C} [ -s \cdot Q s + s \cdot B y] \to \max_{a,C}
\end{equation}
In order to evaluate mathematical expectation in (\ref{exp_fitness_max1}), we start with the first term (quadratic part). Using linearity of the expectation, the covariance formula and symmetricity of matrix $C$, respectively, we calculate the expectation of a quadratic form:
$$
{\mathbb E}(s^T Qs) = {\mathbb E} \left( \sum \limits_{i=1}^n \sum \limits_{j=1}^n q_{ij}s_i s_j \right) = \sum \limits_{i=1}^n \sum \limits_{j=1}^n q_{ij} {\mathbb E} (s_i s_j) =
$$
$$
\sum \limits_{i=1}^n \sum \limits_{j=1}^n q_{ij} (c_{ij} + a_i a_j) = \sum \limits_{i=1}^n \sum \limits_{j=1}^n q_{ij} c_{ji} + \sum \limits_{i=1}^n \sum \limits_{j=1}^n q_{ij}a_i a_j =
$$
$$
\sum \limits_{i=1}^n (Q C)_{i,i} + a^T Q a = Tr(QC) + a^T Q a.
$$
Taking into account that $s$ and $y$ are mutually independent random variables, distributed as $N(a,C)$, expectation of the bilinear term equals $a \cdot B a$. In whole, (\ref{exp_fitness_max1}) is rewritten as
$$
J(a,C) = {\mathbb E}_{a,C} [s \cdot Q s + s \cdot By] = - a \cdot Q a - Tr(QC) + a \cdot B a \to \max_{a,C}.
$$
Gradient of the above function is:
\begin{equation}
\label{vanilla_grad}
\nabla_{(a,C)} J(a,C) = \left( \begin{array}{c} \nabla_a J(a,C) \\ \nabla_C J(a,C) \end{array} \right) = \left( \begin{array}{c} - 2 Qa + Ba \\ - Q \end{array} \right).
\end{equation}
Underline that (\ref{vanilla_grad}) is the {\bf conventional (vanilla)} gradient of $J$. Natural gradient is given by relation (\ref{nat_grad}) which involves the Fisher information matrix.  Multiplication of (\ref{vanilla_grad}) with the inverse of the Fisher information matrix for multivariate Gaussians (\cite{MP}, see also \cite{Beyer}), yields an expression for natural gradient of $J(a,C)$:
$$
\tilde \nabla_{(a,C)} J(a,C) = F^{-1} \nabla_{(a,C)} J(a,C) = \left( \begin{array}{c} C \nabla_a J(a,C) \\ 2 C \nabla_C J(a,C) C \end{array} \right) = \left( \begin{array}{c} C (B - 2 Q)a \\ - 2 C Q C \end{array} \right).
$$
Hence, the natural gradient flow for maximization of $J(a,C)$ is
\begin{eqnarray}
\label{replicator_Gaussian}
\frac{da(t)}{dt} = C(t)(B - 2Q) a(t); \nonumber \\
\frac{dC(t)}{dt} = - 2 C(t) Q C(t).
\end{eqnarray}
Equations (\ref{replicator_Gaussian}) are exactly the system of ODE's reported in Theorem 4. We infer this in the following

\begin{theorem}
Under assumptions 1 and 2, replicator dynamics (\ref{cont_replicator}) are the natural gradient flow for maximization of the mean fitness (\ref{exp_fitness_max1}) on the manifold ${\cal N}(a,C)$.
\end{theorem}

In the next Section we discuss this result in the context of NES and IGO.

\section{Replicator equations as Natural Evolution Strategies}\label{sec:7}

In order to analyze solutions of (\ref{replicator_Gaussian}) first notice that ODE for the covariance matrix does not depend on the mean vector $a(t)$. It is easy to check that the solution is given by $C(t) = (C(0)^{-1} + 2tQ)^{-1}$. Since $Q$ is positive definite, it is evident that $C(t)$ converges to the zero matrix as $t \to \infty$. More precisely, $C(t) \simeq \frac{1}{2t} Q^{-1}$ for a sufficiently large $t$.

On the other hand, convergence of the mean vector depends on eigenvalues of $B - 2Q$. In particular, if all eigenvalues of $B-2Q$ have negative real parts, then the delta distribution at zero is a limiting point for the dynamics (\ref{replicator_Gaussian}) (i.e. $C(t) \to 0$ and $a(t) \to 0$ as $t \to \infty$). Asymptotic properties of solutions of (\ref{replicator_Gaussian}) are analyzed in \cite{CHR}.

Underline that convergence of (\ref{replicator_Gaussian}) is slow, of an order $1/t$. Such a slow convergence might seem surprising at the first glance, but is clear in view of previous considerations.

Rewrite optimization problem (\ref{nat_grad_opt}) for the particular case of maximization of quadratic-bilinear fitness on the manifold ${\cal N}(a,C)$
\begin{eqnarray}
\label{nat_grad_opt_Gauss1}
J(a+\delta a, C + \delta C) - J(a,C) = \nonumber \\
- 2 \delta a \cdot Q a - \delta a \cdot Q \delta a - Tr(Q \cdot \delta C) + 2 \delta a \cdot B a + \delta a \cdot B \delta a \to \max_{(\delta a,\delta C)}
\end{eqnarray}
\begin{eqnarray}
\label{nat_grad_opt_Gauss2}
\mbox{  so that    } I_{KL}(N(a + \delta a,C + \delta C) \, || \, N(a,C)) = \nonumber \\
\frac{1}{2} \left[ \log \frac{\det C}{\det (C+\delta C)}- n + \delta a \cdot C^{-1} \delta a + Tr(I + C \cdot \delta C) \right] = \varepsilon.
\end{eqnarray}
In the above equation we have used an explicit formula for the K-L divergence between two Gaussian random variables, see \cite{MP}. Now, as the particular case of Theorem 3, we have the following

\begin{proposition}
Discretization of replicator equations (\ref{replicator_Gaussian}) yields the gradient ascent algorithm for the problem (\ref{nat_grad_opt_Gauss1})-(\ref{nat_grad_opt_Gauss2}). This algorithm updates parameters $a$ and $C$ of the Gaussian distribution in the direction $(\delta a,\delta C)$ that achieves the largest increase of the mean fitness (\ref{nat_grad_opt_Gauss1}) for a constrained K-L divergence between distributions $N(a+\delta a,C + \delta C)$ and $N(a,C)$.
\end{proposition}

In the light of Proposition 1, slow convergence of the replicator dynamics is expected. Discretization of (\ref{replicator_Gaussian}) yields algorithm with an adaptive step. This algorithm performs slowly at those points on the statistical manifold where the measure changes rapidly. In other words, the step is adapted to the curvature of statistical manifold.

Another insight follows from interpretation of the K-L divergence $I_{KL}(p(\theta + \delta \theta) \, || \, p(\theta))$ as a {\it change of diversity} when the parameter $\theta$ is varied by $\delta \theta$. In view of this observation Proposition 1 can be reformulated in the following way: {\it Replicator equations (\ref{replicator_Gaussian}) maximize the mean fitness in the population, with a constrained loss of diversity}. Notice, however, that preservation of diversity occurs greedily at each step, hence, there is no guarantee that the population will preserve the highest possible diversity for a given value of the mean fitness. Nevertheless, the population that evolves by replicator equations realizes a compromise between maximization of its mean fitness and preservation of the diversity. For further considerations on this aspect of natural gradient algorithms we refer to \cite{OAAH}.

\section{Acceleration of evolutionary dynamics: Fitness shaping in evolutionary games}\label{sec:8}

We have seen that replicator dynamics maximize the mean fitness in the population. However, convergence is slow, since these dynamics follow the natural gradient. In the black-box optimization, fast convergence of algorithms is usually the most important issue. To that end, optimization methods based on the natural gradient are modified in order to ensure faster convergence. These modifications include sampling of several points at each step and evaluation of their fitness. (Note that evaluations of fitness can be computationally expensive.) Those points that have higher fitness are assigned with higher weights. Based on these weights, the probability distribution is updated at the next step. This introduces ideas of genetic algorithms into NES: recombinations, mutations and survival of the fittest.

The idea of sampling a small number of points at each step can be extracted from the more general principle: {\it Monotonically growing transformations of the fitness function do not modify the optimization algorithm}. In other words, one can replace the objective function $f(x|\theta)$ with $W[f(x|\theta)]$, where $W$ is a certain monotonously growing function. Then the iterative procedure for maximization of $W$ will be the same as one for $f$. Techniques of choosing transformation $W$ are called {\it fitness shaping} in optimization, see \cite{WSPS}.

The function $W_f(x|\theta) \equiv W[f(x|\theta)]$ is usually chosen in such a way to perform sampling of a small number of points at each step and leave only those that have higher fitness. It is essential that $W$ depends on the current probability distribution (i.e. on the set of parameters $\theta$). Hence, it is appropriate to denote it $W(x|\theta^t)$, thus emphasizing that $W$ is an {\it adaptive} transformation of $f$. In whole, fitness shaping of $f$ yields a new optimization problem
\begin{equation}
\label{fitness_shaping}
{\mathbb E}[W_f(x|\theta^t)] = \int W_f(x|\theta) p(x|\theta) dx \to \max_\theta.
\end{equation}
We refer to \cite{Beyer,OAAH} for more details on fitness shaping methods and exact choices of the transformation functions $W_f$.

Transformed fitness function (\ref{fitness_shaping}) can be maximized using the natural gradient algorithm. The corresponding flow is given by ODE:
$$
\frac{d \theta}{dt} = F^{-1} \nabla_\theta {\mathbb E} [W_f(x|\theta)] \bigg \vert_{\theta = \theta(t)}.
$$
The vanilla gradient in the above integral is evaluated as
$$
\nabla_\theta {\mathbb E} [W_f(x|\theta)] = \int W_f(x | \theta) \nabla_\theta p(x|\theta) dx.
$$
Hence, the gradient operator {\it does not} act on $W_f$.

As an example, we will briefly discuss CMA ES algorithm with the fitness shaping, as proposed by Arnold in \cite{Arnold}. Without going into details, we only mention that Arnold proposes assigning weights to the fittest points, combined with isotropic Gaussian mutations with carefully chosen mutation strength. It has been shown that Arnold's algorithm ensures the exponential convergence for minimization of the linear-quadratic function. Moreover, Beyer (\cite[Section 4.3]{Beyer}) derived ODE's that are natural gradient flow for Arnold's modification of CMA ES.

Fitness shaping can be included into rules of an evolutionary game. Underline that in this case the invariance of Gaussian distributions is not automatic as it was in the standard setup (where it was ensured by the bilinear-quadratic fitness landscape). Here, we assume that parameters of the Gaussian distribution are updated at each step, based on sampled traits. In addition, only those traits that have highest fitness values contribute to the update. In such a way one can consider discretization of the replicator dynamics, enriched with Arnold's sampling and mutations.

Fitness shaping in evolutionary games have a clear interpretation: although in theory there may exist a continuum of different traits, only a finite number of them is actually realized. They are sampled from the current population state at each step. Besides, only the fittest among them produce an offspring. In addition, one can also introduce mutations of traits into the dynamics. Recombinations and survival of the fittest greatly improve convergence rate. However, there is a price: this obviously happens at the expense of diversity.

Adaptation of Beyer's derivations to the setup of an evolutionary games yields the following modification of replicator equations (\ref{replicator_Gaussian}):
\begin{eqnarray}
\frac{da}{dt} = \frac{C(t)(B - 2Q)a(t)}{\sigma_f(a(t),C(t))} \nonumber \\
\frac{dC}{dt} = - 2 \frac{C(t)QC(t)}{\sigma_f(a(t),C(t))}, \nonumber
\end{eqnarray}
where $\sigma_f(a,C) = \sqrt{(a^T(t)(B^T - 2Q) C (B - 2Q) a(t) + Tr[(QC)^2]}.$
Analysis of the asymptotic dynamics of the above system of ODE's shows that their solutions also converge towards an equilibrium point $(a,C) = (0,0)$ (provided that all eigenvalues of $B - 2 Q$ have negative real parts). However, in contrast to replicator dynamics (\ref{replicator_Gaussian}), this convergence is exponentially fast, i.e. $C(t) \simeq Q^{-1} e^{-\gamma t}$.

Concluding this Section, we stress that fitness shaping can yield a great acceleration of the evolutionary dynamics. This does not come as a surprise: a game in which only the fittest survive exhibits an exponentially fast convergence. At the same time, it causes a rapid loss of the diversity.

Discussion in this Section might be significant, since the idea of designing evolutionary games for multi-agent reinforcement learning is attracting a growing attention (see, for instance, \cite{BTHK,Hennes}). In those cases when fast convergence is a priority, one should borrow fitness shaping methods from the black-box optimization and implement them into evolutionary dynamics. On the other hand, algorithms based on replicator dynamics (possibly enriched with mutations, but without fitness shaping) converge slowly, but are more appealing as a compromise between exploration and exploitation.

\section{Conclusions and outlook}\label{sec:9}

There is a variety of approaches to evolutionary games. For games with a finite strategy set, evolutionary dynamics emerge from the assumption that each individual follows a set of reasonably simple {\it learning rules} (also called {\it revision protocols}, \cite{Sandholm}). For instance, suppose that each player selects a random individual from the (finite) population, and compares their fitness values. If the selected individual has higher fitness, then the player adopts his strategy with a probability proportional to the difference of fitness values. This learning rule generates the replicator dynamics in games with a finite strategy set and linear fitness landscape, see \cite{Schlag}. This can be thought of as "microfoundations" of EGT, that illustrate how individual decision making leads to the collective evolution.

On the other hand, one might argue that results exposed in the present paper can be seen as "macrofoundations" of EGT that unveil the collective ratio: {\it The population as a whole tends to maximize its mean fitness at each step, with a constrained loss of diversity.}

The last assertion is valid for evolutionary dynamics on two particular statistical manifolds: (a) manifold ${\cal P}(A)$ of categorical distributions (\cite{Harper1}); and (b) manifold ${\cal N}(a,C)$ of multidimensional Gaussian measures (the present paper). It is unclear (and remains a challenging open question) if restrictions of dynamics (\ref{cont_replicator}) to other statistical manifolds are also natural gradient flows.

%On the other hand, one might consider an alternative way of introducing evolutionary dynamics by saying the following: natural gradient flow that maximizes the mean fitness on a given invariant family of probability measures is called replicator dynamics. However, it is unclear if such a definition of replicator dynamics is equivalent to the equation (\ref{cont_replicator}) in general (i.e. for statistical manifolds different from categorical and Gaussian distributions).

The present paper has to a great extent been inspired by parallel developments in EGT and evolutionary algorithms of optimization. The link between these two fields is based on the concepts of NES and IGO. To our best knowledge, relations between EGT and black-box optimization have not been recognized and studied before.

Information Geometry provides a natural unifying framework for the study of evolutionary games, as it allows a principled approach which is valid for games with a finite strategy set, as well as for games with a continuous trait space. In the upcoming studies we will introduce new evolutionary games and investigate information-geometric aspects of evolutionary dynamics on some other statistical manifolds, such as Cauchy and wrapped Cauchy distributions.

Possibly the most promising are potential applications to Artificial Intelligence (AI) and Deep Learning (DL). As emphasized above, there is a growing interest in analogies between evolutionary games and multi-agent reinforcement learning (MARL), \cite{BTHK,Hennes}. On the other hand, natural gradient has been introduced in \cite{Amari1,AN} with a view on possible applications in DL. Since then, natural gradient has been employed in many algorithms in AI and DL, see \cite{Martens} for the conceptual and fairly practical survey. The present paper might contribute to applications of evolutionary games with a continuous trait space to MARL. For instance, AI algorithms that use natural gradient can be implemented through evolutionary games. In particular, deep neural networks and generative adversarial networks might be trained through the evolutionary dynamics.

Natural gradient flows do not exhibit fast convergence. Although this looks like a drawback from the point of view of black-box optimization, it is actually an advantage in some situations. Algorithms that exploit natural gradient (and, hence, replicator equations) can be of a great interest for designing AI algorithms with an inherently incorporated compromise between exploration and exploitation.

One peculiarity of EGT that makes it an exciting field of research is that it can be justified and interpreted in many different ways. It has been founded as a branch of Game Theory about 50 years ago, by introducing game-theoretic paradigms into Theoretical Biology. It has been later recognized that models of EGT might present even greater significance for social sciences, and even Law and Philosophy. Furthermore, it has been pointed out that the evolutionary dynamics can be seen as a learning process, similar to Bayesian inference, and this observation led to intriguing hypotheses in Cognitive Science (roughly speaking, the brain tends to maximize novelty, while keeping surprise constrained, \cite{FA}). Recently, information-geometric aspects of evolutionary games have inspired new conceptual approaches in Phylogeny (\cite{CGTS}).

From the theoretical perspective, EGT still poses new challenges to different fields of mathematics, including dynamical systems, optimization, distributed control, information geometry and Shanon theory of information.

Meanwhile, EGT is frequently questioned and criticized for the explanatory irrelevance of its findings, \cite{Alexander}. This criticism might be addressed by answering that EGT itself is a game, invented in order to amuse students and motivate them to think about concepts of evolution, strategy, algorithm, collective learning, etc. However, if evolutionary games enter the realm of AI more concretely in the upcoming years, then one should admit that this game is getting pretty serious.

\section*{Acknowledgements}
The author acknowledges the partial support of the Ministry of Science of Montenegro, project titled: "Mathematical Modeling of Collective Motions: Distributed Optimization and Control" (abbreviation: MMCM-DUO)





% \end{thebibliography}

\end{document}